\documentclass[a4paper]{jpconf}
\usepackage{graphicx}

\newcommand{\AddrAHEP}{AHEP Group, Institut de F\'{i}sica Corpuscular --
  C.S.I.C./Universitat de Val\`{e}ncia, Parc Cientific de Paterna.
  C/Catedratico Jos\'e Beltr\'an, 2 E-46980 Paterna (Val\`{e}ncia) - SPAIN}

\newcommand{\Cali}{Universidad Santiago de Cali, Campus
  Pampalinda, Calle 5  No. 6200, 760001, Santiago de Cali, Colombia}

\newcommand{\Cinvestav}{Departamento de F\'{\i}sica, Centro de
  Investigaci{\'o}n y de Estudios Avanzados del IPN\\ Apdo. Postal
  14-740 07000 Mexico, DF, Mexico}
  
\newcommand{\iu}{{i\mkern1mu}}
    
\begin{document}
\title{New limits on neutrino magnetic moments from low energy neutrino data}

\author{B C Ca\~nas$^1$, O G Miranda$^1$, A Parada$^2$, M T\'ortola$^3$ and J W F Valle$^3$}

\address{$^1$~\Cinvestav}
\address{$^2$~\Cali}
\address{$^3$~\AddrAHEP}


\ead{bcorduz@fis.cinvestav.mx}

\begin{abstract} 

  Here we give a brief review on the current bounds on the general
  Majorana transition neutrino magnetic moments (TNMM) which cover also
  the conventional neutrino magnetic moments (NMM). Leptonic CP phases
  play a key role in constraining TNMMs. While the Borexino experiment
  is the most sensitive to the TNMM magnitudes, one needs
  complementary information from reactor and accelerator experiments
  in order to probe the complex CP phases.

\end{abstract}

\section{Introduction}\vskip .2cm

The study of neutrino electromagnetic properties is of great
importance, since it could open a new window to investigate physics
beyond the Standard Model.  
Though there are various types of electromagnetic
properties~\cite{Giunti:2015gga,Giunti:2014ixa}, such as a neutrino
charge
radius~\cite{Bardeen:1972vi,Lee:1973fw,Barranco:2007ea,Kosmas:2015sqa}
or a neutrino milli-charge~\cite{Marciano:1977wx,Papoulias:2015iga},
here we concentrate on the case of
TNMM~\cite{Schechter:1981hw,Kayser:1982br,Bell:2007nu}.
These are appealing for several reasons. For instance, like
neutrinoless double beta decay~\cite{Schechter:1981bd}, TNMMs can shed
light on the fundamental issue of the Dirac or Majorana nature of
neutrinos~\cite{Schechter:1981hw,Kayser:1982br,Bell:2007nu}.
They are also expected to be calculable in a gauge theory, their
finite values given in terms of fundamental neutrino properties, such
as masses and mixing parameters, in addition to other genuine new
physics parameters such as new messenger particle masses.
The main constraints on neutrino electromagnetic properties come from
reactor neutrino studies~\cite{Li:2002pn,Deniz:2009mu} as well as from
solar neutrino data~\cite{Beacom:1999wx,Grimus:2002vb}.
There are, moreover, many proposals aiming to improve the current
bounds, one of them using a megacurie $^{51}$Cr neutrino source and a
large liquid Xenon detector~\cite{Coloma:2014hka}. 
Based on the analysis reported in~\cite{Canas:2015yoa}, we summarize
some of the most relevant constraints on TNMM and discuss their
dependence on the CP phases.

We include different types of neutrino data samples, such as the most
recent data from the TEXONO reactor experiment~\cite{Deniz:2009mu}, as
well as the latest results from the Borexino
experiment~\cite{Bellini:2011rx}.
Data from the reactor experiments Krasnoyarsk~\cite{Vidyakin:1992nf},
Rovno~\cite{Derbin:1993wy} and MUNU~\cite{Daraktchieva:2005kn} along
with the accelerator experiments LAMPF~\cite{Allen:1992qe} and
LSND~\cite{Auerbach:2001wg} are also included.
In addition, we take into account the updated values of the neutrino
mixing parameters as determined in global oscillation
fits~\cite{Forero:2014bxa}, including the value of $\theta_{13}$
implied by Daya-Bay~\cite{An:2013zwz} and RENO~\cite{Ahn:2012nd}
reactor data, as well as accelerator data~\cite{Abe:2013hdq}. Besides,
we stress on the role played by the, yet unknown, leptonic CP
violating phases.

\section{Neutrino magnetic moments}\vskip .2cm
\label{sect:NMM}
The interaction between Majorana neutrinos and the electromagnetic
field is described by the general effective
Hamiltonian~\cite{Schechter:1981hw}
\begin{equation}
\textit{H}^{M}_{em} = -\frac14\nu^{T}_{L}C^{-1}\lambda\sigma^{\alpha\beta}\nu_{L}F_{\alpha\beta} + h.c. ,
\end{equation}
with $\lambda = \mu - id$ a complex antisymmetric matrix in generation
space, implying that $\mu^{T}=-\mu$ and $d^{T}=-d$ are pure
imaginary. Therefore, we need six real parameters to describe the
Majorana NMM.  The Majorana NMM matrix can be written in the flavor
(mass) basis, $ \lambda$ ($\tilde{\lambda}$), as follows
\begin{equation}
\lambda =
\left(\begin{array}{ccc}
0 & \Lambda_{\tau} & -\Lambda_{\mu} \\
-\Lambda_{\tau} & 0 & \Lambda_{e} \\
\Lambda_{\mu} & -\Lambda_{e} & 0 \end{array}\right) , \qquad
\tilde{\lambda} =
\left(\begin{array}{ccc}
0 & \Lambda_{3} & -\Lambda_{2} \\
-\Lambda_{3} & 0 & \Lambda_{1} \\
\Lambda_{2} & -\Lambda_{1} & 0 \end{array}\right) .
\label{Eq:nmm-matrix}
\end{equation}
Here we have defined
$\lambda_{\alpha\beta} =
\varepsilon_{\alpha\beta\gamma}\Lambda_{\gamma}$, through the complex
parameters:
$\Lambda_{\alpha}=|\Lambda_{\alpha}|e^{\iu\zeta_{\alpha}}$,
$\Lambda_{i}=|\Lambda_{i}|e^{\iu\zeta_{i}}$.
Having described our theoretical framework, we now discuss the
relation of these observables with the parameters measured in current
neutrino experiments.
%
%
For neutrino-electron scattering, the differential cross section for
the NMM contribution will be given by
\begin{equation}\label{eq:xsec_mm}
\left(\frac{d\sigma}{dT}\right)_{{em}}  = \frac{\pi \alpha^{2}}{m^{2}_{e}\mu^{2}_{B}}\left(\frac{1}{T}-\frac{1}{E_{\nu}}\right){\mu_{\nu}}^{2} ,
\end{equation}
where $\mu_{\nu}$ is an effective magnetic moment accounting for the
NMM contribution to the scattering process. It is defined in terms of
the components of the NMM matrix in Eq.~(\ref{Eq:nmm-matrix}). In the
flavor basis, this parameter can be written as ~\cite{Grimus:2002vb}
\begin{equation}\label{Eq:meff_F}
({\mu_{\nu}^{F}})^{2} = a^{\dag}_{-}\lambda^{\dag}\lambda a_{-} + a^{\dag}_{+}\lambda\lambda^{\dag}a_{+},
\end{equation}
where we have denoted the negative and positive helicity neutrino
amplitudes by $a_{-}$\, and \,$a_{+}$, respectively.  The flavour and
mass neutrino basis are connected through the neutrino mixing matrix U
\begin{equation}
\tilde{a}_{-} = U^{\dag}a_{-},\qquad \tilde{a}_{+}=U^{T}a_{+},\qquad \tilde{\lambda}=U^{T}\lambda U,
\end{equation}
such that the effective NMM in the mass basis is given by
\begin{equation}
({\mu_{\nu}^{M}})^{2} = \tilde{a}^{\dag}_{-}\tilde{\lambda}^{\dag}\tilde{\lambda} \tilde{a}_{-} + \tilde{a}^{\dag}_{+}\tilde{\lambda}\tilde{\lambda}^{\dag}\tilde{a}_{+}.
\end{equation}
Notice that there are six complex phases in the effective NMM
parameter: $\zeta_1$, $\zeta_2$ and $\zeta_3$ from the NMM matrix;
$\delta$ and two-Majorana phases from the leptonic mixing matrix.
However, as noticed in Ref.~\cite{Grimus:2000tq}, it is clear that
only three of these six complex phases are independent.  To carry out
our analysis~\cite{Canas:2015yoa}, we choose the Dirac CP phase
$\delta$, and the two relative phases, $\xi_2 = \zeta_3 - \zeta_1$ and
$\xi_3 = \zeta_2 - \zeta_1$.


\begin{figure}[t!]
\centering
\includegraphics[width=0.85\linewidth]{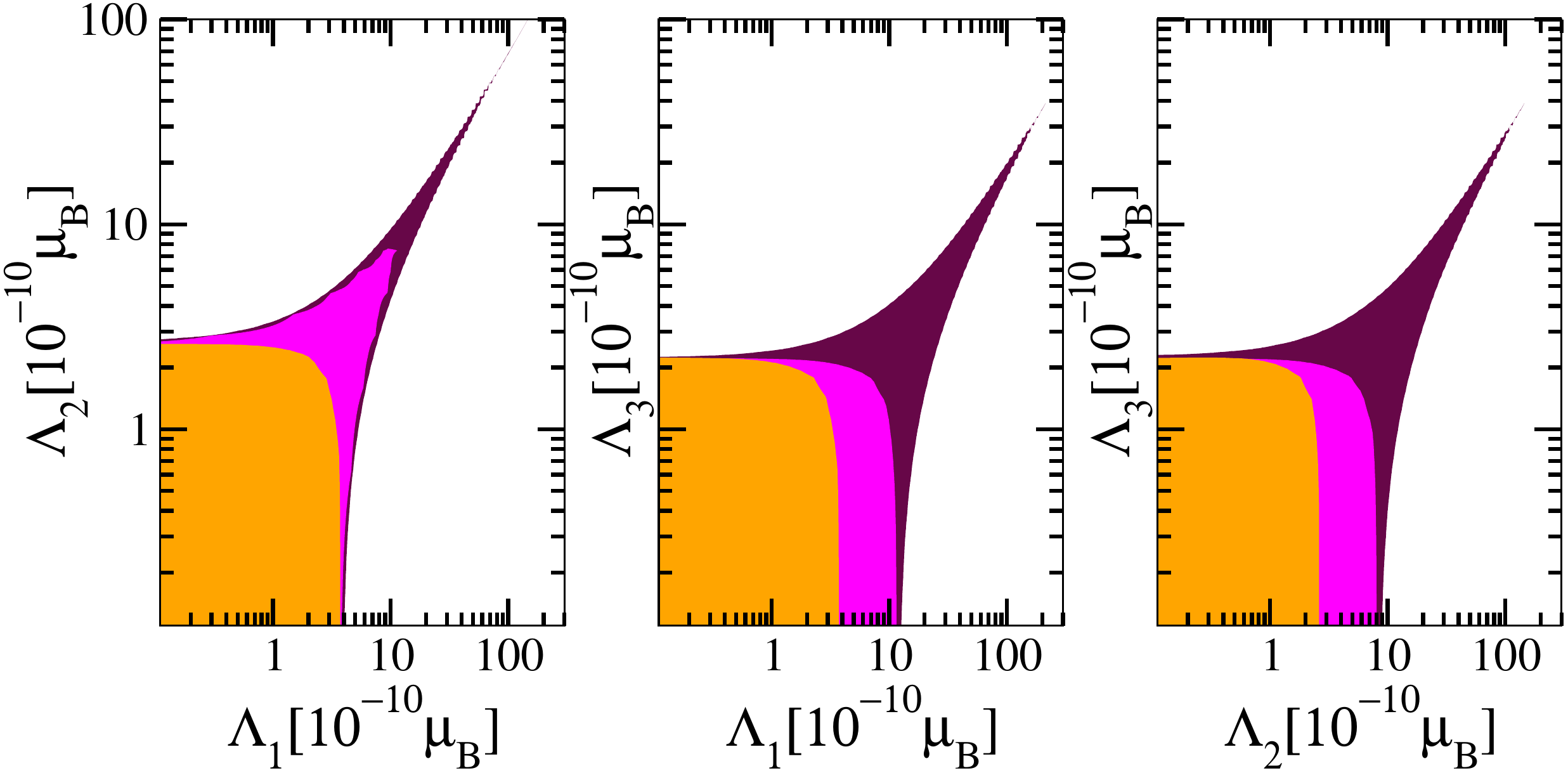}
\caption{Allowed regions for the TNMM, at 90 \% CL, obtained from 
TEXONO data. 
The two-dimensional regions ($|\Lambda_i|$, $|\Lambda_j|$) have
been computed by marginalizing over the third component,
$|\Lambda_k|$.
For the purple (outer) region we have fixed the phases to be
$\delta = 0$ and $\xi_2 = \xi_3 = 0$ (see Eq.~(\ref{eq:mureac2})), while 
in the magenta (middle) region we have set 
$\delta = 3\pi/2$ and $\xi_2 = \xi_3 = 0$.  The orange (inner) region 
corresponds to $\delta=3\pi/2$,
$\xi_2=0$ and $\xi_3 =\pi/2$.  }
\label{fig.reactor.phase}
\end{figure}

In reactor experiments, we initially have only an electron
antineutrino flux and, therefore, $a^1_{{+}}=1$ will be the only
nonzero entry.  With this initial condition, we get from
Eq.~(\ref{Eq:meff_F}) the following expression for the effective
Majorana NMM for reactor antineutrino experiments in the flavor basis:
\begin{equation}\label{mureac_flav}
(\mu^{F}_{R})^{2} = |\Lambda_{\mu}|^{2} + |\Lambda_{\tau}|^{2} .
\end{equation}
On the other hand, for the mass basis, we have the expression
\begin{eqnarray}\label{eq:mureac}
(\mu^{M}_{R})^{2} &=& {|{\bf \Lambda}|^{2}} - s^{2}_{12}c^{2}_{13}|\Lambda_{2}|^{2} - c^{2}_{12}c^{2}_{13}|\Lambda_{1}|^{2} - s^{2}_{13}|\Lambda_{3}|^{2}\\\nonumber 
&-& 2s_{12}c_{12}c^{2}_{13}|\Lambda_{1}||\Lambda_{2}|\cos\delta_{12}
- 2c_{12}c_{13}s_{13}|\Lambda_{1}||\Lambda_{3}|\cos\delta_{13}\\\nonumber
&-& 2s_{12}c_{13}s_{13}|\Lambda_{2}||\Lambda_{3}|\cos\delta_{23},
\end{eqnarray}
with $c_{ij} = \cos\theta_{ij}$, $s_{ij} = \sin\theta_{ij}$. The
phases in this equation depend on the three independent CP phases
already mentioned: $\delta_{12}= \xi_3$,
$\delta_{23}= \xi_2 - \delta$, and
$\delta_{13}= \delta_{12}-\delta_{23}$.
The dependence on these CP phases is very interesting and adds an
extra complexity to the interpretation of the experimental
constraints. For instance, in the particular case where all the
independent phases vanish, i.e,
$\delta_{12}=\delta_{23}=\delta_{13}=0$, the effective Majorana NMM in
Eq.~(\ref{eq:mureac}) is given by
\begin{eqnarray}\label{eq:mureac2}
(\mu^{M}_{R})^{2} &=& {|{\bf \Lambda}|^{2}} - (c_{12}c_{13}|\Lambda_{1}|+s_{12}c_{12}c_{13}|\Lambda_{2}|
                   +  s_{13}|\Lambda_{3}|)^{2}.
\end{eqnarray}
It is easy to notice that if we also impose the conditions
\begin{equation}\label{eq:mureac0}
 |\Lambda_{1}| = c_{12}c_{13}|{\bf \Lambda}|, \quad |\Lambda_{2}| = s_{12}c_{13}|{\bf \Lambda}|, \quad |\Lambda_{3}| = s_{13}|{\bf \Lambda}|,
\end{equation}
$(\mu^{M}_{R})^{2}$ cancels exactly and, hence, in this case reactor
experiments would not be sensitive to this parameter. This is
illustrated in Fig.~\ref{fig.reactor.phase}.

A similar situation will appear in accelerator experiments such as the
LAMPF~\cite{Allen:1992qe} and LSND~\cite{Auerbach:2001wg},
where a dependence on the CP phases will
appear~\cite{Canas:2015yoa}.

On the other hand, in solar neutrino experiments like
Borexino~\cite{Bellini:2011rx}, the effective NMM parameter in the
mass basis is given by
\begin{equation}\label{eq:nmm_sun}
(\mu^{M}_{\rm{sol}})^{2} = |\mathbf{\Lambda}|^{2} -
  c^{2}_{13}|\Lambda_{2}|^{2} + (c^{2}_{13}-1)|\Lambda_{3}|^{2} +
  c^{2}_{13}P^{2\nu}_{e1}(|\Lambda_{2}|^{2}-|\Lambda_{1}|^{2}),
\end{equation}
where we have defined the effective two-neutrino oscillation
probabilities as $P^{2\nu}_{e1,e2}$. Due to the loss of coherence, the
effective NMM measured from the solar neutrino flux is independent of
any phase, a fact already noticed in~\cite{Grimus:2002vb}. Notice also
that the analysis presented here takes into account the non-zero value
of the reactor angle $\theta_{13}$~\cite{Canas:2015yoa}.

\section{Analysis of the neutrino data}\vskip .2cm

We perform a combined analysis of the experimental data in order to
get constraints for the three different TNMM components $\Lambda_i$.
In order to carry out the statistical analysis we use the following
$\chi^2$ function:
\begin{equation}\label{eq:chi2}
\chi^{2}=\sum^{N_{bin}}_{i=1}
\left(\frac{O^i- N^i(\mu)} {\Delta_i}\right)^{2},
\end{equation}
where $O^i$ and $N^i$ are the observed number of events and the
predicted number of events in the presence of an effective NMM, $\mu$,
at the $i$-th bin, respectively.  Here $\Delta_i$ is the statistical
error for each bin.  In our analysis, we have included the
experimental results reported by Krasnoyarsk~\cite{Vidyakin:1992nf},
Rovno~\cite{Derbin:1993wy}, MUNU~\cite{Daraktchieva:2005kn}, and
TEXONO~\cite{Deniz:2009mu} reactor experiments.  We have also included
the experimental data reported by the LAMPF~\cite{Allen:1992qe} and
LSND~\cite{Auerbach:2001wg} collaborations, as well as 
the most recent measurements of the Beryllium solar neutrino flux
reported in Ref.~\cite{Bellini:2011rx} by Borexino.

We perform a complete analysis taking into account the role of the
CP phases in the reactor and accelerator data.
For the particular case of reactor neutrinos, we have carried out a
statistical analysis of TEXONO data~\cite{Deniz:2009mu} taking
different values of the complex phases of $\Lambda_i$, $\zeta_i$, and
taking all nonzero TNMM amplitudes. The result of this analysis is
shown in Fig.~\ref{fig.reactor.phase}.  Notice that the resulting
restrictions on the TNMM $|\Lambda_1|$ and $|\Lambda_2|$ depend on the
chosen CP phase combinations.

\begin{figure}[t]
\centering
\includegraphics[width=0.75\linewidth]{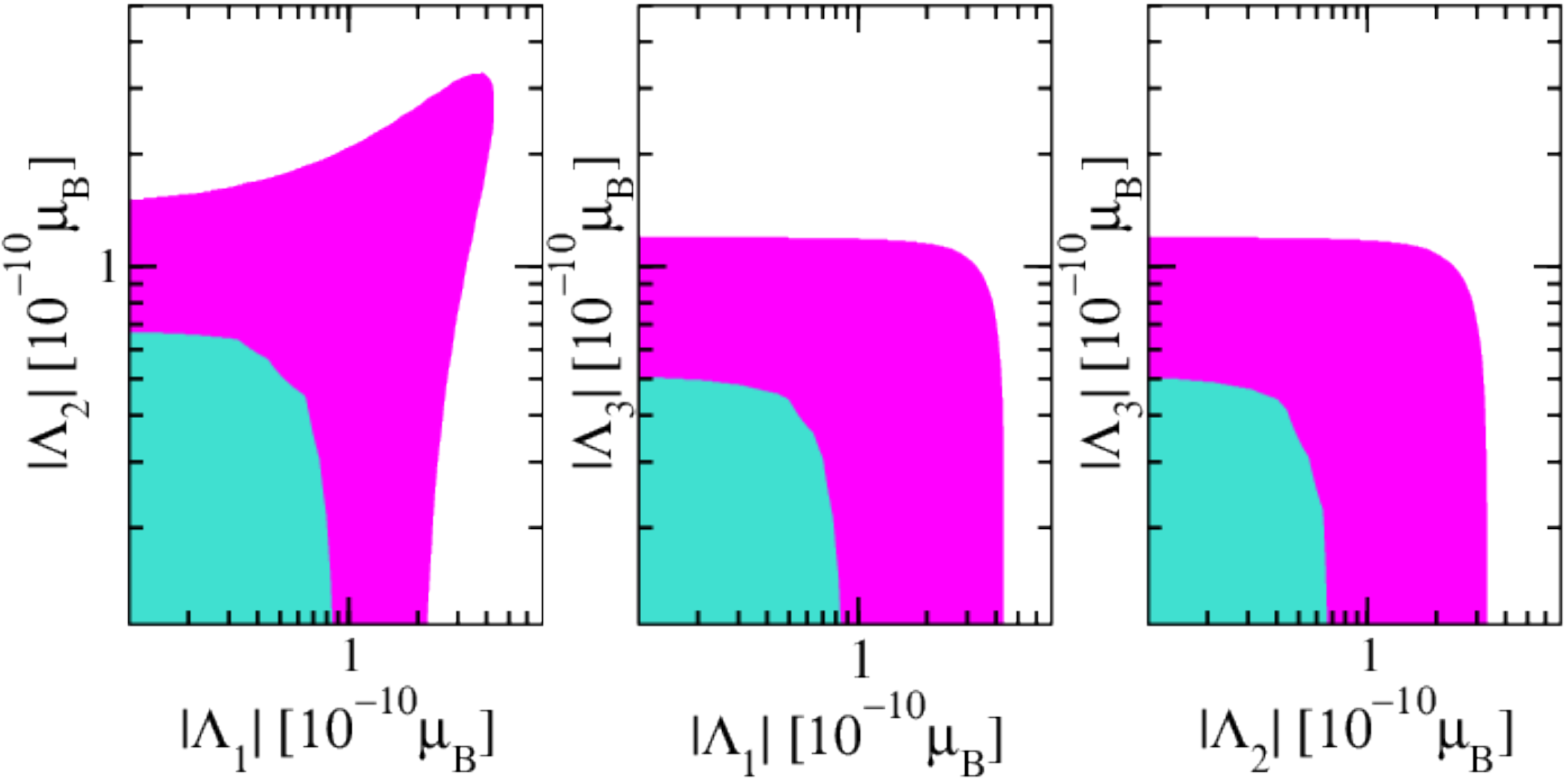}
\caption{Allowed regions, at 90 \% CL, for the
  TNMM~\cite{Canas:2015yoa}. The magenta region shows the constraints
  from reactor and accelerator data when all the phases vanish,
  except for $\delta = 3\pi/2$. The turquoise zone shows the
  corresponding bounds from the Borexino data that are phase-independent.}
\label{fig:combined-new}
\end{figure}

  Finally, we carried out a combined analysis~\cite{Canas:2015yoa}
  of all the reactor and accelerator data for a particular choice of
  phases ($\delta=3\pi / 2$ and $\xi_i = 0$) and compared it with the
  corresponding $\chi^2$ analysis obtained from the Borexino data.
The results, shown in Fig.~\ref{fig:combined-new}, illustrate how
Borexino~\cite{Bellini:2011rx} is more sensitive in constraining the
magnitude of the TNMM. We stress that the Borexino effective NMM
depends only on the square magnitudes of these TNMM and hence, its
constraints are almost the same as those in the
one-parameter-at-a-time analysis.  On the other hand, future reactor
and accelerator experiments are the only ones that could give
information on individual TNMMs as well as on the Majorana phases
discussed here, an information inaccessible to the Borexino
experiment.

\section{Conclusions}\vskip .2cm

In this short review, we have discussed the current status of the
bounds on the transition neutrino magnetic moments.  These parameters
are very important, because they encode the Majorana CP phases present
both in the mixing matrix and in the NMM matrix.  The conventional
neutrino magnetic moment emerges as a particular effective case.
The Borexino solar experiment plays a key role in constraining the
electromagnetic neutrino properties due to the low energies (below 1
MeV) which are probed as well as its robust statistics. Indeed, it
provides the most stringent constraints on the absolute magnitude of
the transition magnetic moments. However, the Borexino experiment can
not probe the Majorana phases, due to the incoherent nature of the
solar neutrino flux. Although less sensitive to the absolute value of
the transition magnetic moment strengths, reactor and accelerator
experiments provide the only chance to obtain a hint of the complex CP
phases, as illustrated in Fig.~\ref{fig.reactor.phase}.

\section*{Acknowledgements}

Work supported by MINECO grants FPA2014-58183-P, Multidark CSD2009-
00064 and SEV-2014-0398 (MINECO); by EPLANET, by the CONACyT grant
166639 (Mexico) and the PROMETEOII/2014/084 grant from Generalitat
Valenciana.
M.~T\'ortola is also supported by a Ramon y Cajal contract of the
Spanish MINECO.
The work of A.Parada was partially supported by Universidad Santiago
de Cali under grant 935-621114-031.

\section*{References}

\end{document}